\documentclass{SPAICE}
\usepackage{float}
\def\authorEmail{freja.thoresen@proton.me}
\def\AuthorShort{F. Thoresen et al.}

\author[1,2]{Freja Thoresen\thanks{Corresponding author. E-Mail: \authorEmail}}
\author[2]{Aidan Cowley}
\author[2]{Romeo Haak}
\author[2]{Jonas Lewe}
\author[2]{Clara Moriceau}
\author[2]{Piotr Knapczyk}
\author[2,3]{Victoria S. Engelschiøn}

\affil[1]{Directorate of Human and Robotics Exploration, European Space Agency (ESA), European Space Research and Technology Centre (ESTEC), Noordwijk, Netherlands}

\affil[2]{Directorate of Human and Robotics Exploration, European Space Agency (ESA), European Astronaut Centre (EAC), Cologne, Germany}

\affil[3]{The Natural History Museum, University of Oslo, Oslo, Norway}

\title{Breccia and basalt classification of thin sections of Apollo rocks with deep learning\thanks{Accepted at the ESA SPAICE 2026 Conference.}}

\begin{document}

\maketitle

\begin{abstract} 
Human exploration of the Moon is expected to resume in the next decade, following the last such activities during the Apollo programme. One of the major objectives of returning to the Moon is to continue retrieving geological samples, with a focus on collecting high-quality specimens to maximize scientific return. Tools that assist astronauts in making informed decisions about sample collection activities can maximize the scientific value of future lunar missions. A lunar rock classifier can provide astronauts with the necessary information to analyze lunar rock samples and augment in-situ value identification. In this paper, we introduce a framework for classifying rock types in thin sections of lunar rocks.
We leverage the vast collection of petrographic thin-section images from the Apollo missions, captured under plane-polarized light (PPL), cross-polarised light (XPL), and reflected light at varying magnifications. Contrastive learning is applied to analyze these images and extract meaningful features. A pre-trained Inception-ResNet-v2 network is fine-tuned using the SimCLR loss function to effectively extract features from thin-section images of Apollo rocks. A binary classifier trained with transfer learning from the fine-tuned Inception-ResNet-v2 achieves 93.51\% ($\pm$3.17) image-level accuracy and 98.83\% ($\pm$0.95) sample-level accuracy in separating breccias from basalts.
\end{abstract}

\section{Introduction}
The analysis of lunar samples is crucial to understanding the formation, evolution, and resource potential of the Moon \cite{NewViewsOfTheMoon}. Hence, rock sample collection is one of the important scientific objectives of returning to the Moon. Information on rock type or specific mineralogy may help study the lunar geology and can be used for decision-making, e.g., if the material is suitable for a particular experiment or return to Earth \cite{Crawford2012}. 

The vast collection of Apollo rocks has been studied extensively. A common geological method is petrographic thin sections, where rock specimens are polished to a thickness of 30 $\mu$m \cite{Heiken1991, Meyer2005, allton1989catalog}. These are then examined in an optical microscope by rotation under differing light conditions to determine the minerals and textures within the rocks, known as petrography \cite{UoAGeology}. While effective, this method is time-consuming and requires expertise; subsequently,  there is a need for automated or semi-automated systems for analyzing lunar rocks in order to maximize the scientific output of lunar missions. In recent years, machine learning has shown to be a powerful tool for automating the analysis of thin-section images \cite{Liu2024}. Deep learning, especially convolutional neural networks (CNN), has impressive performance in image classification and segmentation tasks \cite{Prince2023, DBLP:journals/corr/abs-2001-04074}.

We introduce a method for classifying rock textures in thin sections of lunar breccias and basalts and establish a framework for additional analysis including segmentation and clustering, aiding tools that help astronauts identify rocks of interest and guide targeted sampling to maximise scientific return. In summary, our contributions are:
\begin{itemize}
    \item Curate a dataset of thin-section images of Apollo lunar rocks from the NASA PDS and LPI databases.
    \item Fine-tune a pre-trained Inception-ResNet-v2 network using SimCLR contrastive learning to extract meaningful petrographic features from lunar thin sections.
    \item Train a binary breccia/basalt classifier via transfer learning from the SimCLR-fine-tuned network, achieving 93.51\% ($\pm$3.17) accuracy at the image level and 98.83\% ($\pm$0.95) at the sample level.
    \item Establish a foundation for further analysis of lunar thin sections, including segmentation and clustering, using the fine-tuned feature extraction network.
\end{itemize}

\subsection{Lunar Rock Types}
The Moon mainly consists of  the igneous rocks anorthosite and basalt, and breccias, which are aggregates of broken rock fragments cemented together. Anorthosite is a calcium-rich rock that largely makes up the ancient crust of the Moon \cite{windley1970anorthosites}. 
Lunar basalts are the product of the partial melting of the lunar mantle, primarily found in the Mare regions of the Moon \cite{Heiken1991}. They are low in silica and contain more iron and less total silicon ($\text{SiO}_2$) than terrestrial basalts. Lunar basalts typically comprise pyroxene, plagioclase, olivine, and metal oxides such as ilmenite, armalcolite, and spinel \cite{Heiken1991}. 
Breccias are composite rocks formed by impacts, comprising clasts of older rock bound together by a finer matrix \cite{Wood1974, Heiken1991}. Most lunar breccias formed in the highlands under impact heat and pressure \cite{Heiken1991}. 

Over the years, a collection of labeled thin-section images from the Apollo missions has been created \cite{Apollo11, Apollo12, Apollo14, Apollo15, Apollo16, Apollo17, PostApollo}. The thin sections are made by cutting, smoothing, and polishing rocks to a 30$\mu$m thickness and are typically imaged under various lighting conditions (plane-polarized light (PPL), cross-polarized light (XPL), and reflected light) \cite{Meyer2005}. The number of thin sections per sample varies significantly, ranging from 2 to 70, depending on the extent of prior studies. Each thin section has multiple images, often at varying magnifications (2.5-10x) and focusing on different regions of the section.

\begin{figure}[H]
    \centering
    \includegraphics[width=0.5\linewidth]{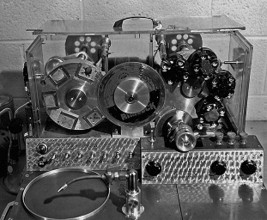}
    \caption{In 1965, the United States Geological Survey (USGS) developed a prototype for a semi-automatic machine to cut and prepare thin sections of rocks \cite{cary1965}. This innovative device was specifically designed for the Apollo Applications Program and Advanced Lunar Programs Groups, aiming to streamline the analysis of lunar samples.}
    \label{fig:device}
\end{figure}

While thin sections of lunar rocks were created when they were returned and analyzed on Earth, an interesting prospect is to make the thin sections in a lunar laboratory. In \cite{Foucher2021}, designs for an automated rock thin section device are suggested, which may efficiently cut future rock samples in a space environment. In \cref{fig:device}, a concept of a semi-automatic machine is shown, which was developed in 1965. Development of tools and payloads to support astronauts with in-situ science analysis capability represents a new paradigm for exploration and will be increasingly important in future missions beyond LEO - ultimately, a level of laboratory capability will be required as Moon and even Mars missions make terrestrial return logistics challenging.

\subsection{Machine Learning for Petrographic Thin Section Analysis}
Machine learning has been applied to segment, classify, and cluster petrographic thin sections across a range of geological contexts. Segmentation approaches include using multi-angle polarization images with a Mask R-CNN trained via transfer learning from ResNet34 \cite{Liu2022}, and a model combining discrete convolutional filters with random forest classifiers to segment carbonate thin sections into mineral and porosity classes \cite{Rubo2019}. For classification, ResNet50 and ResNet101 achieve 90.24\% and 91.63\% accuracy on thin-section images \cite{Guojian2021}, while a concatenated CNN classifying 13 rock types from PPL and XPL images achieves 89.97\% accuracy with a Kappa coefficient of 0.86 \cite{Su2020}. Together, these studies demonstrate the effectiveness of deep learning for both pixel-level segmentation and image-level classification of petrographic thin sections.

\section{Data}
\begin{figure}[H]
    \centering
    \includegraphics[width=0.99\linewidth]{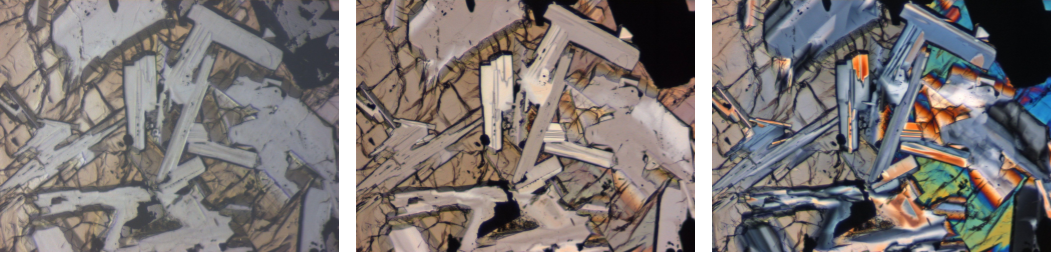}
    \caption{Example of a thin section under different polarized light conditions: (left) Reflected light, (center) plane-polarized light (PPL), and (right) cross-polarised light (XPL).}
    \label{fig:light}
\end{figure}

Two databases are used: the NASA PDS database \cite{NASA_PDS3_ApolloLunarSamples} and the LPI Lunar Sample Atlas \cite{LPI_LunarSampleAtlas}. The PDS database contains digitally scanned Apollo photographs under three lighting conditions (PPL, XPL, and reflected light), as seen in \cref{fig:light}. Only thin-section images were used, selecting the full-resolution JPEG format, with label files providing sample metadata. This yields 365 samples and 5203 images.

The LPI dataset similarly contains digitally scanned Apollo photographs. Selecting thin sections of Apollo rocks yields 605 samples and 17066 images; samples 10072A and 10072 were excluded as no labeled data was found.

Both datasets were integrated with duplicates removed; additional metadata were sourced from the Lunar Sample and Photo Catalog Curator \cite{LunarSampleAndPhotoCatalogCurator}. Images with mislabelling, manual markings, or less than 60\% thin-section coverage were excluded. Black borders and scanning artefacts were cropped from LPI images, and an erroneous image reference in sample 12003 was removed. The overall distribution is in Table 1.  

\subsection{Data Preprocessing}
In the data preprocessing, the dataset is split into training (60\%), validation (20\%), and a test set (20\%). For each split, images of the same sample end up in the same set. 
The images are resized to 299x299x3, and the pixel color values of the images are normalized. The data is augmented by a layer in the training of the network, with augmentations of random vertical/horizontal rotations, random zoom, brightness, saturation, hue, and contrast. The augmentations artificially create more data, making the network more robust to small perturbations, thus preventing overfitting \cite{Perez2017}.

As seen in Table 1, the dataset is imbalanced. Therefore, class frequencies are calculated to generate class weights to counteract potential problems arising from the imbalance. During the training, the class weights are then applied to the dataset, making the less abundant class more relevant, thus improving the imbalance.

\begin{table}[]
\centering
\begin{tabular}{ccc}
\toprule
\textbf{Category} & \textbf{Samples} & \textbf{Images} \\ \midrule
Basalt            & 187              & 4769            \\ 
Breccia           & 326              & 8470            \\ 
Total             & 513              & 13239           \\ \bottomrule
\end{tabular}
\caption{Distribution of samples and images by rock texture category.}
\end{table}

\begin{figure*}[h]
    \centering
    \includegraphics[width=0.99\linewidth]{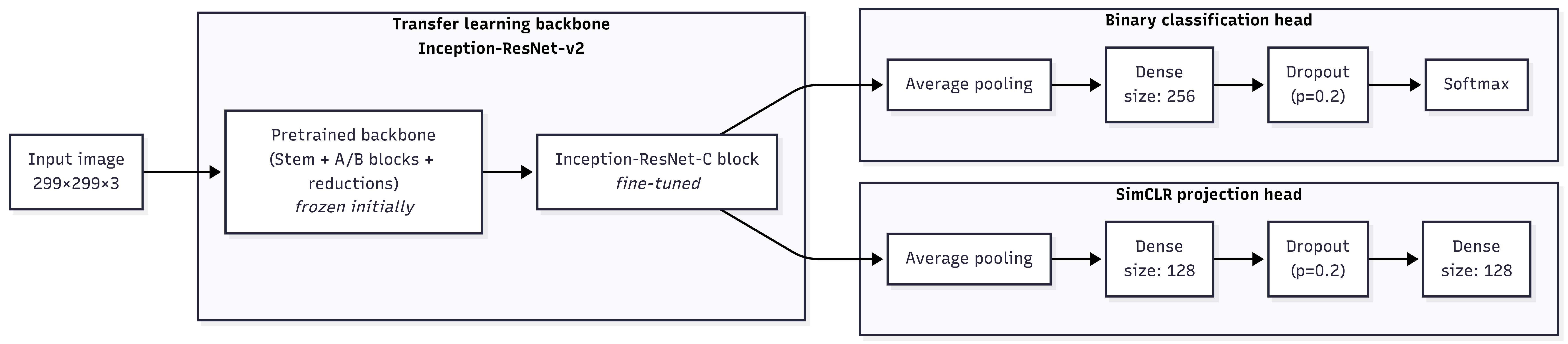}
    \caption{Overview of the network architecture. (Left) Transfer learning backbone based on Inception-ResNet-v2 \cite{Szegedy2016}, taking $299\times299\times3$ input images; the pre-trained stem, Inception-ResNet-A/B blocks and reduction blocks are frozen initially, while the Inception-ResNet-C block is fine-tuned. (Top right) Binary breccia/basalt classification head: average pooling, dense layer (256 units), dropout ($p=0.2$) and softmax. (Bottom right) SimCLR projection head: average pooling, dense layer (128 units), dropout ($p=0.2$) and dense layer (128 units), which is discarded after SimCLR training is complete.}
    \label{fig:network_architecture}
\end{figure*}

\section{Network Architecture}

The methodology employed in this paper will leverage contrastive learning to extract significant representations from images of thin sections of lunar rock. The implementation of contrastive learning will be utilized within a network consisting of a transfer model of convolutional residual connections and a classification head. Subsequently, the derived representations will serve as input for training a binary classifier. In the following sections, these fundamental concepts will be introduced in detail.

\subsection{Transfer Learning with Inception Residual Networks}
The introduction of residual connections accelerated research on neural networks \cite{He2015}. Residual connections enable training of deeper networks and mitigate the vanishing gradient problem by adding the block input to the layer transformation via identity mappings, allowing gradients to flow more effectively during backpropagation. ResNet is a convolutional neural architecture that vastly improved the accuracy of deep neural networks by implementing this approach. 

It was found that deeper convolutional networks with the inception principles proved effective in improving the use of computing resources within the network, hence improving the network while keeping computational resources constant \cite{Szegedy2015}. Due to the success of both inception and ResNet networks, we choose to employ an Inception-ResNet-v2 \cite{Szegedy2016} as the underlying transfer learning model, which is pre-trained on ImageNet. 
 
\cref{fig:network_architecture} shows the structure of the transfer learning model. The first 6 blocks depicted are a group of multiple layers (mostly convolutional) defined in \cite{Szegedy2016}. After the first 6 layers are taken from the Inception-ResNet-v2 network, the network is equipped with classification heads, one for the SimCLR training and another for the binary classifier.

\subsection{Contrastive Learning with SimCLR}
\begin{figure}
    \centering
    \includegraphics[width=0.6\linewidth]{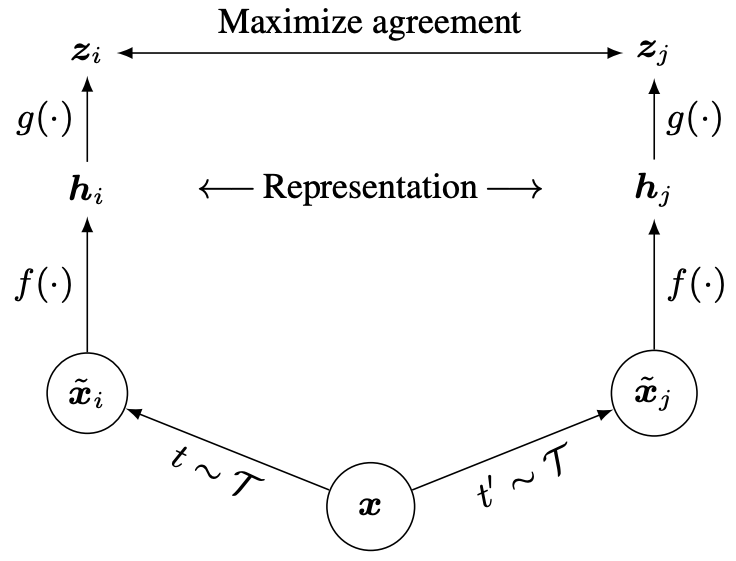}
    \caption{A simple framework for contrastive learning of visual representations from \cite{Chen2020}. }
    \label{fig:simclr}
\end{figure}

Contrastive learning methods, such as a Simple Framework for Contrastive Learning of Visual Representations (SimCLR) \cite{Chen2020}, have successfully learned generalisable features for various image tasks with limited data. SimCLR learns its general representations with the use of data augmentations. The process involves simultaneous optimisations of agreement among distinct transformations of the same image while minimizing concordance among transformations of dissimilar images. In this study, the neural network is trained using SimCLR. The SimCLR approach is in \cref{fig:simclr}. The input images $x$ are augmented into two representations, $\hat{x_i}$ and $\hat{x_j}$. Then, the representations are passed through the neural network, which is, in this paper, the Inception-ResNet-v2 network, resulting in the feature representations $h_i$ and $h_j$. The feature representations are passed through a shallow and wide network, $g(\cdot)$, which then results in the final output of $z_i$ and $z_j$. The loss function then maximizes the agreement between the two representations while minimizing the agreement between representations of other images. After the training is complete, the classification head $g(\cdot)$ is removed, and the transfer model is fine-tuned according to the SimCLR approach and can be used for other classification or clustering tasks. The architecture of the network is in \cref{fig:network_architecture}.

\subsection{Breccia/Basalt Binary Classifier}
As in the contrastive learning training, the training of the breccia/basalt binary classifier is divided into two phases: The first phase freezes the transfer learning model and trains the classification head, and the second phase fine-tunes the combined transfer model and classification head. The network is in \cref{fig:network_architecture}, where the transfer learning model is pre-trained with the SimCLR approach.

During the first phase, the transfer learning model is frozen up till the second block of Inception-ResNet C, as it strikes a good balance between fine-tuning specialized features while not leading to a large entropic capacity and hence a strong tendency to overfit \cite{Chollet2016}. The Adaptive Moment Estimation (Adam) \cite{Kingma2014}, with the AMSgrad version, is used, as it has been shown to be relatively more robust \cite{Reddi2019}. Though the learning rate is adaptive, the Adam optimizer is supplied with an initial learning rate of 0.001.

In the fine-tuning phase, the frozen layers are unfrozen. In this phase, a non-adaptive learning rate optimizer, Stochastic Gradient Descent (SGD), is used, with a low learning rate of 0.0001, to keep the updates small in the final tuning. Both phases are equipped with early stopping features \cite{Prechelt1998} in order not to overtrain. To estimate the performance of the model, a stratified K-fold cross-validation was performed.

\section{Results}
\begin{figure}
    \centering
    \includegraphics[width=0.85\linewidth]{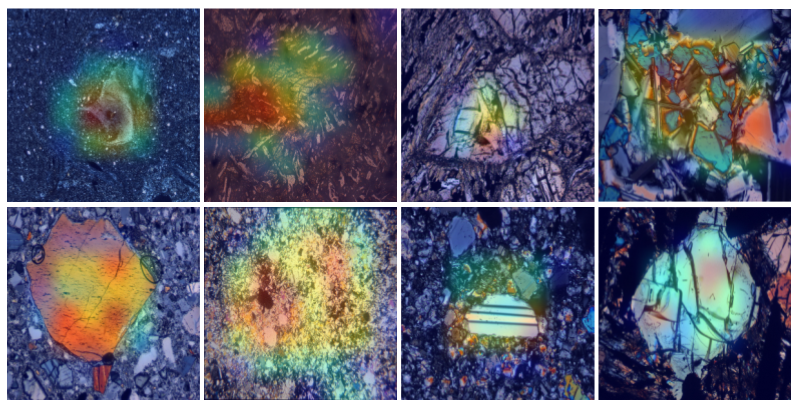}
    \caption{Example of activations of the SimCLR network on images of thin sections of lunar rock. }
    \label{fig:gradcam}
\end{figure}

The SimCLR network was trained for at least 100 epochs, with the option of early stopping, since it has been observed that SimCLR benefits from extended training. \cref{fig:gradcam} illustrates the output of the final layer of Inception-ResNet-v2 trained with SimCLR, which represents the features that will be passed to the binary classifier. It effectively highlights various features, such as the shapes of minerals within a sample.

Since the samples have multiple pictures, the predictions of individual images can be grouped to predict the sample's rock type, similar to the approach of ensemble learning. When the predictions for each sample are combined, it results in a better prediction, as is seen in Table 2. The resulting model proved successful, with an image accuracy of 93.51\% ($\pm$3.17) and a group accuracy of 98.83\% ($\pm$0.95), calculated on the test set. The results demonstrate that convolutional neural networks can effectively classify lunar rock types into their major categories.

\begin{table}[H]
\centering
\begin{tabular}{lcc}
\toprule
\textbf{Level} & \textbf{Accuracy (\%)} & \textbf{Std (\%)} \\ \midrule
Image  & 93.51 & $\pm$3.17 \\
Sample & 98.83 & $\pm$0.95 \\ \bottomrule
\end{tabular}
\caption{Binary breccia/basalt classifier accuracy at the image level and sample level (K-fold cross-validation on the test set).}
\end{table}

\section{Discussion}
The binary breccia/basalt classifier achieves an image-level accuracy of 93.51\% ($\pm$3.17) and a sample-level accuracy of 98.83\% ($\pm$0.95), demonstrating that deep learning can reliably distinguish between the two dominant lunar rock texture classes. The sample-level ensemble outperforms image-level prediction alone, as multiple thin-section images of the same sample capture complementary regions and magnifications, reducing the impact of ambiguous individual frames. The class imbalance (187 basalt, 326 breccia samples) was mitigated with class weights, but a balanced dataset would be preferable. 

The Grad-CAM activations shown in \cref{fig:gradcam} confirm that the SimCLR-fine-tuned network attends to mineralogically meaningful image regions, such as grain boundaries, mineral morphologies, and textural fabrics. It suggests the network has learned petrographically relevant representations, and it validates the feature extractor as a suitable backbone for downstream geological analysis tasks such as segmentation and clustering.

Future work could extend classification to breccia subgroups and anorthosite, and mineral-phase segmentation of pyroxene, plagioclase, and ilmenite using hyperspectral or Raman data \cite{Heiken1991}. The SimCLR-fine-tuned model can be reused as a feature extractor for these tasks as the dataset expands.

\section{Conclusion}
In this study, we investigated the use of SimCLR, a contrastive learning method, for extracting features from lunar thin sections. The feature extraction was effective, enabling a binary breccia/basalt classifier trained via transfer learning to achieve 93.51\% ($\pm$3.17) image-level accuracy and 98.83\% ($\pm$0.95) sample-level accuracy. These results establish a foundation for further analysis of Apollo thin sections, with the fine-tuned feature extraction network directly applicable to segmentation, clustering, and multi-class classification tasks as the dataset expands and more rock-type labels become available.

\section{Code Availability}
All code to reproduce the results in this paper is available in the \href{https://github.com/esa/apollo_rock_thin_section_classifier}{GitHub repository}.

\section{Author Contributions}
\textbf{Freja Thoresen:} Conceptualization, project administration, software, supervision, validation, visualization, writing - original draft, writing - review and editing. 
\textbf{Aidan Cowley:} Conceptualization, project administration, resources, supervision, writing - review and editing. 
\textbf{Romeo Haak:} Conceptualization, data curation, formal analysis, investigation, methodology, software, validation. 
\textbf{Jonas Lewe:} Conceptualization, data curation, formal analysis, investigation, methodology, software. 
\textbf{Clara Moriceau:} Conceptualization, data curation, formal analysis, investigation, methodology, software, writing - review and editing. 
\textbf{Piotr Knapczyk:} Conceptualization, data curation, formal analysis, investigation, methodology, software. 
\textbf{Victoria S. Engelschiøn:} Conceptualization, data curation, formal analysis, investigation, methodology, software, writing - review and editing.

\begin{acknowledgments}
We extend our sincere acknowledgements to the co-authors Victoria S. Engelschiøn, Piotr Knapczyk, Jonas Lewe, Clara Moriceau, and Romeo Haak. The research presented in this paper was done during their time as integral members of the Spaceship EAC group at the European Astronaut Centre, part of ESA. Acknowledgements to NASA and LPI for providing the vast dataset of images of the thin sections of lunar rocks.
\end{acknowledgments}

\printbibliography
\addcontentsline{toc}{section}{References}

\end{document}